\documentclass[superscriptaddress,prd,twocolumn,nofootinbib,secnumarabic,floatfix,nofootinbib,aps]{revtex4-1}
\usepackage{amsmath,amssymb,hyperref}
\usepackage{graphicx,epsfig}
\usepackage{slashed}
\usepackage{float,xcolor}
\usepackage[utf8]{inputenc} 
\usepackage[title]{appendix}
\usepackage{subcaption} 
\usepackage{caption}
\usepackage{comment}
\usepackage{ulem}
\usepackage{graphicx}
\usepackage{epstopdf}

\def\beq{\begin{equation}}
\def\eeq#1{\label{#1}\end{equation}}
\def\eeqn{\end{equation}}
\def\beqa{\begin{eqnarray}}
\def\eeqa#1{\label{#1}\end{eqnarray}}
\def\eeqan{\end{eqnarray}}

\newcommand{\centeron}[2]{{\setbox0=\hbox{#1}\setbox1=\hbox{#2}\ifdim
\wd1>\wd0\kern.5\wd1\kern-.5\wd0\fi \copy0
\kern-.5\wd0\kern-.5\wd1\copy1\ifdim\wd0>\wd1
                                   \kern.5\wd0\kern-.5\wd1\fi}}
\newcommand{\ltap}{\>\centeron{\raise.35ex\hbox{$<$}}
                           {\lower.65ex\hbox{$\sim$}}\>}
\newcommand{\gtap}{\>\centeron{\raise.35ex\hbox{$>$}}
                           {\lower.65ex\hbox{$\sim$}}\>}

\begin{document} 

{\hfill FERMILAB-PUB-18-668-A-PPD-T}

\title{Proton Fixed-Target Scintillation Experiment to Search for Minicharged Particles}

\author{Kevin J. Kelly}
\email{kkelly12@fnal.gov}
\affiliation{Theoretical Physics Department, Fermilab, P.O. Box 500, Batavia, IL 60510, USA}
\author{Yu-Dai Tsai}
\email{ytsai@fnal.gov}
\email{\\}
\affiliation{Fermilab, Fermi National Accelerator Laboratory, Batavia, IL 60510, USA}
\affiliation{University of Chicago, Kavli Institute for Cosmological Physics, Chicago, IL 60637}

\date{\today}

\begin{abstract}

We propose a low-cost and movable setup to probe minicharged particles (or milli-charged particles) using high-intensity proton fixed-target facilities. This proposal, FerMINI, consists of a milliQan-type detector, requiring multi-coincident (nominally, triple-coincident) scintillation signatures within a small time window, located downstream of the proton target of a neutrino experiment. During the collisions of a large number of protons on the target, intense minicharged particle beams may be produced via meson photo-decays and Drell-Yan production. We take advantage of the high statistics, shielding, and potential neutrino-detector-related background reduction to search for minicharged particles in two potential sites: the MINOS near detector hall and the proposed DUNE near detector hall, both at Fermilab. We also explore several alternative designs, including the modifications of the nominal detector to increase signal yield, and combining this detector technology with existing and planned neutrino detectors to better search for minicharged particles. The CERN SPS beam and associated experimental structure also provide a similar alternative. FerMINI can achieve unprecedented sensitivity for minicharged particles in the MeV to few GeV regime with fractional charge $\varepsilon=Q_{\chi}/e $ between $10^{-4}$ (potentially saturating the detector limitation) and $10^{-1}$.

\end{abstract}

\maketitle

\section{Introduction}

The quantization of electric charge, as currently observed in nature, has been one of the longest-standing mysteries in particle physics. The Standard Model (SM) $U(1)$ hypercharge group in principle allows arbitrarily small charges, yet experiments so far suggest that electric charge has a fundamental unit. This has inspired the concept of Dirac quantization~\cite{Dirac:1931kp} and motivated several considerations of Grand Unification Theorie (GUT) (see, e.g. Refs.~\cite{Pati:1973uk,Georgi:1974sy}), among other theoretical studies. The discovery of particles with electric charge magnitude less than the smallest quark charges, namely minicharged particles (MCP) would be a major paradigm shift in this endeavor. MCP have been studied and searched for on various fronts (see, e.g., Refs.~\cite{Dobroliubov:1989mr,Prinz:1998ua,Davidson:2000hf,Prinz:2001qz,Golowich1987,Babu:1993yh,Gninenko:2006fi,CMS:2012xi,Agnese:2014vxh,Haas:2014dda,Ball:2016zrp,Alvis:2018yte,Magill:2018tbb})\footnote{Although charge quantization could never truly be ruled out, even with the detection of such particles, their discovery would weaken the concept. It would also test the predictions of theories related to charge quantization (e.g. \cite{Pati:1973uk,Georgi:1974sy,Shiu:2013wxa}).}.

Given that MCP interact feebly with the SM particles through their small electric charge, they are also a potential solution to another well-established mystery of particle physics: dark matter. MCP may be low-energy consequences of fermions in the dark sector \cite{Alexander:2016aln} that couple to the Standard Model via a massless dark photon through kinetic mixing~\cite{Holdom:1985ag}, and the dark sector particles may constitute the relic abundance of dark matter. The theories and signatures of dark sector and dark photon have been heavily explored (see, e.g., Refs.~\cite{Brahm:1989jh,Boehm:2003hm,Pospelov:2007mp,Bjorken2009,Batell2009,deNiverville:2011it,Izaguirre:2013uxa,Batell:2014mga,Kahn:2014sra,Dobrescu:2014ita,Coloma:2015pih,deNiverville:2016rqh,Pospelov:2017kep, Magill:2018jla}). Most recently, MCP as dark matter has been proposed to explain the anomalous 21 cm hydrogen absorption signal reported by the Experiment to Detect the Global Epoch of Reionization Signature (EDGES) collaboration~\cite{Bowman:2018yin,Barkana:2018lgd, Munoz2018}. However, orthogonal constraints have been explored, and it has been demonstrated that the favored MCP candidates to explain the EDGES result do not comprise the entirety of the observed relic dark matter abundance (see, e.g., Refs.~\cite{Berlin:2018sjs,Barkana:2018qrx, Slatyer:2018aqg}). The favored range of masses of these MCP is below roughly a hundred MeV, which is a region that could be explored in proton fixed-target experiments.

Probes of MCP and other weakly-interacting MeV-GeV particles has been under intense study, due in large part to the fact that many dark matter and dark sector hypotheses fall into these categories. Additionally, experimental techniques to probe this region have matured significantly~\cite{Alexander:2016aln}. The most sensitive laboratory-based probes of MCP are threefold:
\begin{itemize}
	\setlength\itemsep{-0.35em}
	\item Collider Probes
	\item Electron Fixed-Target Experiments
	\item Proton Fixed-Target and Neutrino Experiments 
\end{itemize}
Both the Tevatron and Large Hadron Collider (LHC) have provided constraints on MCP for the first category~\cite{Davidson:2000hf,CMS:2012xi}. Additionally, a dedicated experiment (milliQan) was specifically proposed to occupy the CMS P5 site to search for MCP~\cite{Haas:2014dda,Ball:2016zrp}.
The Beijing Electron–Positron Collider could also provide sensitivity to the MCP particles \cite{Liu:2018jdi}.
Electron fixed-target experiments have been historically the most sensitive searches for MCP below 100 MeV. The dedicated SLAC MCP experiment \cite{Prinz:1998ua,Prinz:2001qz} still provides leading sensitivity for MCP in this range. Several proposed electron-fixed target experiments (e.g. LDMX \cite{Berlin:2018bsc} and NA64 \cite{Gninenko:2018ter}) can further improve the sensitivity of MCP, but the mass reach would be limited by the beam energy. Finally, using neutrino experiments and protons-on-fixed-targets to study MCP has been long proposed~\cite{Golowich1987,Babu:1993yh,Gninenko:2006fi}, but a dedicated analysis considering all the current and proposed near-future experiments has only been done recently \cite{Magill:2018tbb}, followed by the study based on reactor neutrino experiment for the lower mass MCP \cite{Singh:2018von}.
 
Here, we propose a Fermilab-based experiment to probe minicharged particles, FerMINI, that combines the techniques of dedicated searches at SLAC~\cite{Prinz:1998ua,Prinz:2001qz} and the LHC~\cite{Haas:2014dda} with the advantages of neutrino facility sites~\cite{Magill:2018tbb}. 
We utilize the intense proton beams, for example, the existing Neutrinos at the Main Injector (NuMI) beamline and the future Long-Baseline Neutrino Facility (LBNF) beamlines, and place more than one (the nominal design is 3) groups of scintillator arrays downstream of the intense beam, shielded from strong electromagnetic radiation.
The signature for MCP is the detection of one or a few photoelectrons (PE) produced when the particle traverses the scintillator, causing small ionizations and producing photons collected by the photomultiplier tubes (PMT). We require such a detection in contiguous scintillator-PMT sets in each of the three detector groups in order to greatly reduce background. The two sites we explore are the existing Main Injector Neutrino Oscillation Search (MINOS) near detector hall and the proposed Deep Underground Neutrino Experiment (DUNE) near detector hall. We show the potential reach of such setups in Figs.~\ref{fig:MoneyPlot_NuMI} and~\ref{fig:MoneyPlot_DUNE}, respectively, and see that these can provide leading sensitivity to MCP searches for masses in the range of 10 MeV to 5 GeV. 

The nominal FerMINI setup substantially benefits from the large fluxes of MCP from the intense proton collisions. We will also discuss new ideas to combine the MCP detector with the neutrino detectors in Section \ref{sec:Alternatives}.
FerMINI serves as an example to demonstrate that the proton-fixed target facilities could be natural habitats for the dedicated low-cost detectors, including milliQan (mostly proposed for LHC recently), searching for weakly interacting and long-lived particles.

Note that, in the analysis, we limit our attention to the minimal theoretical assumption that the MCP we are searching for are simply fermions with small $U(1)_Y$ hypercharges with masses between MeV-GeV (if the minicharged particle is a scalar instead, the sensitivity is largely similar). The model and constraints do not rely on the existence of dark photons nor assumptions of MCP abundance and velocity distributions in the local galaxy, but the bounds we derive certainly serve as conservative constraints to the MCP-related dark matter and dark sector scenarios. Interestingly, finding MCP without an accompanying massless dark photon would have implications on not only GUT theories (again see, e.g., Refs.~\cite{Pati:1973uk,Georgi:1974sy}) but also string compactifications and quantum gravity \cite{Shiu:2013wxa}. The subtleties between different MCP scenarios are further explored in Ref.~\cite{Tsai:2018_Thesis,MCP_Proposal}. One can also use this setup to search for dark sector particles that couple to the SM through a massive vector field, as demonstrated in Refs.~\cite{Izaguirre:2015eya,milliQan:slides}. Outside of probing MCPs and light dark matter scenarios, one can also utilize this proposal to probe the electric dipole moment of a heavy neutrino, as was proposed utilizing the milliQan facility \cite{Sher:2017wya}. We leave the detailed analysis on these fronts to a future study \cite{MCP_Proposal}.

\section{Production}\label{sec:production}

We consider minicharged particles $\chi$ with electric charge $Q_\chi$ and define $\varepsilon \equiv Q_\chi/e$. For $m_\chi < 10$ GeV, existing constraints bound $\varepsilon \lesssim 10^{-1}$, and even stronger constraints exist for $m_\chi \lesssim 100$ MeV. In proton fixed-target experiments, minicharged particles are produced via neutral meson decays and Drell-Yan processes, discussed below:

\textbf{Meson Decays:} We consider the following meson decays to the millicharged particle $\chi$:
\begin{itemize}
	\setlength\itemsep{-0.25em}
\item $\pi^0 \to \gamma\chi\bar{\chi}$  ($m_{\pi^0} = 135$ MeV)
\item $\eta \to \gamma\chi\bar{\chi}$ ($m_\eta = 548$ MeV)
\item $J/\psi \to \chi\bar{\chi}$ ($m_{J/\psi} = 3.1$ GeV)
\item $\Upsilon \to \chi\bar{\chi}$ ($m_\Upsilon = 9.4$ GeV)
\end{itemize}
When produced in proton-proton collisions, each of these mesons $\mathfrak{m}$ may decay into millicharged particles with masses up to $m_\mathfrak{m}/2$.

For $\mathfrak{m} = \pi^0, \eta$, the decay proceeds similar to that of $\mathfrak{m} \to \gamma e^+ e^-$. We may write the total number of $\chi$ produced via these decays as 
\begin{equation}
N_{\chi} \simeq 2 c_{\mathfrak{m}} \mathrm{Br}(\mathfrak{m}\rightarrow \gamma\gamma) \varepsilon^2\alpha_{EM} N_\mathrm{POT} \times I^{(3)}\left(\frac{M^2_{\chi}}{m^2_{\mathfrak{m}}}\right)\, .	
\end{equation}
Here, $c_{\mathfrak{m}}$ is the number of meson $\mathfrak{m}$ produced per proton-on-target (POT, total number $N_\mathrm{POT}$) in the target hall, $\varepsilon^2 \alpha_{EM}$ is proportional to the $\gamma \bar{\chi} \chi$ coupling, and $I^{(3)}(x)$ characterizes the three-body decay\footnote{Note that with $\varepsilon = 1$, the product $\mathrm{Br}(\pi^0 \to\gamma\gamma) I^{(3)} (m_e^2/m_{\pi^0}^2)$ roughly reproduces the observed branching fraction of $\pi^0\to\gamma e^+ e^-$.} $\mathfrak{m} \to \gamma\chi\bar{\chi}$,
\begin{eqnarray}
&&I^{(3)}(x) = \frac{2}{3\pi}\int_{4x}^1 dz \sqrt{1-\frac{4x}{z}} \frac{1-z}{z^2}\times \nonumber \\
&&\left(12x^3+6x^2(3z-2) + x(5z-2)(z-1)+z(z-1)^2\right). \nonumber\\
\end{eqnarray}
We find, using {\tt PYTHIA8}~\cite{Sjostrand:2014zea}, $c_{\pi^0} \simeq 4.5$ and $c_\eta \simeq 0.5$ for $120$ GeV protons on target.

The $J/\psi$ and $\Upsilon$ mesons may decay directly via $\mathfrak{m} \to \chi \bar{\chi}$, and 
\begin{equation}
N_\chi \simeq 2 c_{\mathfrak{m}} Br(\mathfrak{m} \rightarrow e^+ e^-) \varepsilon^2 N_\mathrm{POT} \times I^{(2)}\left(\frac{M^2_{\chi}}{m^2_{\mathfrak{m}}},\frac{m^2_{e}}{m^2_{\mathfrak{m}}}\right),
\end{equation}
where
\begin{equation}
I^{(2)}(x,y) = \frac{\left(1 + 2x\right)\sqrt{1-4x}}{(1+2y)\sqrt{1-4y}}.
\end{equation}
The amount of $J/\psi$ or $\Upsilon$ produced per POT is far smaller than that of $\pi^0$ and $\eta$, though their production is important for larger $m_\chi$. We find that $c_{J/\psi} \simeq 4.4\times 10^{-5}$ and\footnote{Since $c_\Upsilon$ is too small to generate $\Upsilon$ particles via a Monte Carlo generator, we assume the flux of $\chi$ from this decay is similar to that from $J/\psi$ decay, simply extending the maximum allowed $m_\chi$ up to $m_\Upsilon/2$ (instead of $m_{J/\psi}/2$), and scaled by the fraction $c_\Upsilon/c_{J/\psi}$. This scaling fraction is determined in Ref.~\cite{Magill:2018tbb} and the reference therein.} $c_\Upsilon \simeq 2.5 \times 10^{-9}$.

\begin{figure}[!htbp]
\centering
\includegraphics[width=0.49\textwidth]{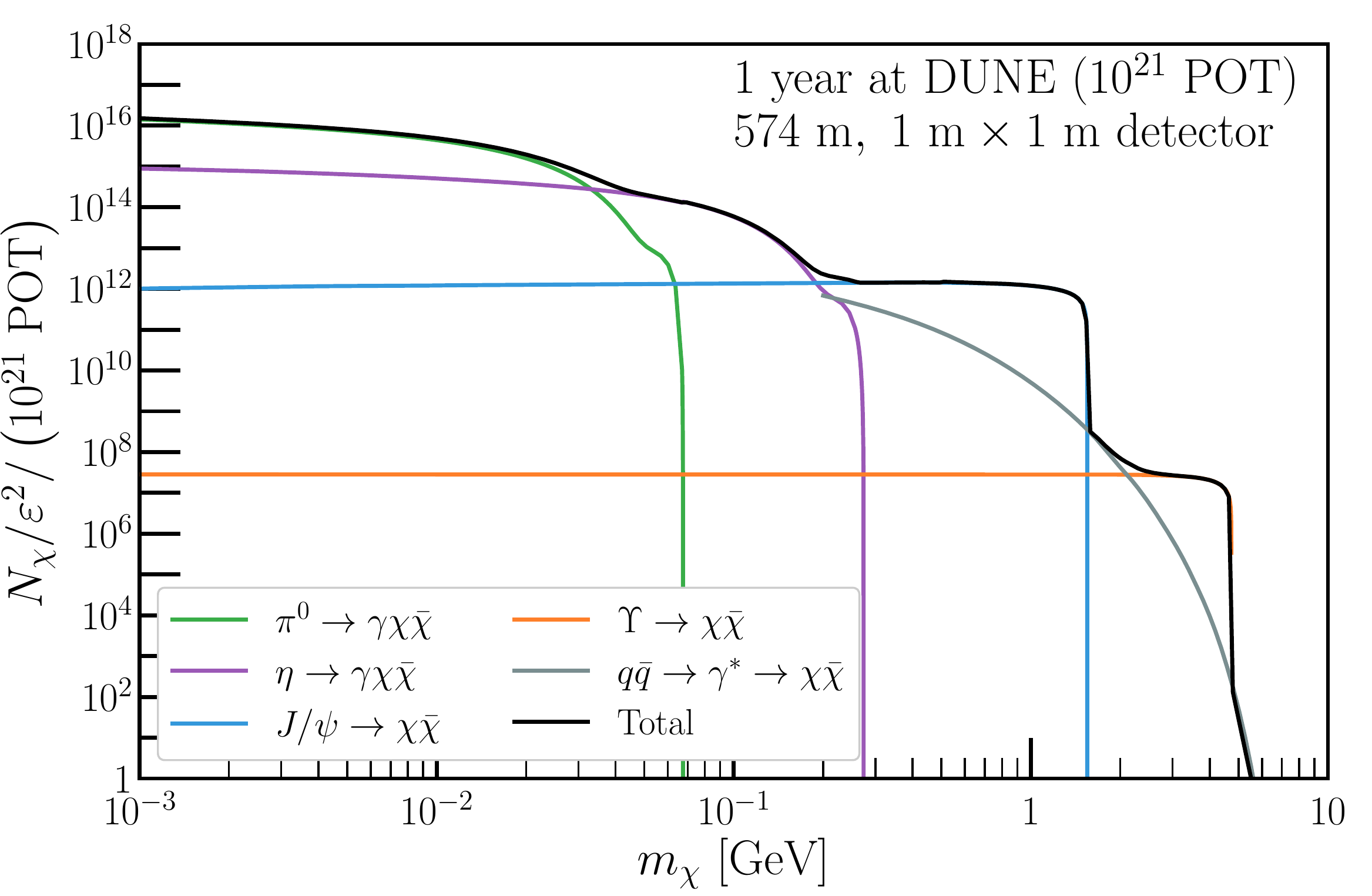}
\caption{Expected number of $\chi$ particles to reach a perpendicular \mbox{1 m $\times$ 1 m} detection area a distance of $574$ m from a 120 GeV proton beam hitting a target, in agreement with the DUNE near detector hall. We assume 1 year of live-trigger time for the accumulated events, corresponding to $10^{21}$ POT. Different colors show the production of $\chi$ via meson decays and Drell-Yan production, as described in the text.}
\label{fig:ChiProduction}
\end{figure}

\textbf{Drell-Yan Production:} We also take into account the Drell-Yan production, $q\bar{q} \to \chi\bar{\chi}$ using MSTW parton distribution functions \cite{Martin:2009iq}. For the 120 GeV protons on target for the NuMI and DUNE beams, this process is subdominant to meson-production, except for when $m_\chi > m_\Upsilon/2$ (see the gray curve in Fig.~\ref{fig:ChiProduction}).

While most of the $\chi$ particles produced via meson decay and Drell-Yan production are forward-going, not all pass through the $1$ m$^2$ cross-section of the detector at a distance of $\sim 500-1000$ m away. We take this into account during our simulation and find that $\mathcal{O}(10^{-4})$ of $\chi$ produced reach the detector.

\section{Nominal Setup $\&$ Signature}
\label{sec:Setup}

Now we focus on the FerMINI experimental design and signature. As mentioned above, the basic idea is to place a milliQan-type detector downstream of a proton beam to detect the MCP flux from the collisions of protons on a fixed target.
In order to accumulate a large flux of MCP, we propose placing the MCP detector downstream of a beam of a neutrino facility, specifically the existing NuMI or the upcoming LBNF beamline at Fermilab. We provide some detail of the two options here. First, we propose placing the detector in the MINOS near detector hall (1040 m downstream of the NuMI target hall and roughly 100 m underground), which is shielded by the absorber (840 m downstream) and 200 meters of rocks between the absorber and the MINOS near detector hall~\cite{Adamson:2015dkw}. One interesting feature of the NuMI beam setup is that about 87 $\%$ of the protons interact at the main target and 13 $\%$ of the protons interact in the absorbor.
This means that there is a secondary beam production of MCP at the absorber, and the MCP produced there would suffer less geometric suppression than those produced at the main target.

The implementation at the LBNF/DUNE\footnote{In our simulations using {\tt PYTHIA8}~\cite{Sjostrand:2014zea} we assume that the LBNF beamline will consist of $120$ GeV protons on target, however we do not expect that an $80$ GeV or $100$ GeV beam would generate significantly different distributions of $N_\chi$ or their resulting energies, for the sake of our analysis.} site~\cite{DUNECollaboration2015} is largely similar. We propose to place the detector in the near detector hall $\sim 570$ meters downstream of the LBNF target hall complex (roughly 60 meters underground).
The shielding from radiation for the LBNF/DUNE setup is also naturally provided by the absorber ($\sim$ 220 meters from the target) followed by muon alcove and roughly 300 meters of rocks. Most protons would interact in the main target hall so the secondary production is negligible in the LBNF/DUNE configuration.

The nominal detector design of FerMINI directly follows \cite{Haas:2014dda,Ball:2016zrp}, which consists of three stacks of scintillator arrays coupled to PMTs for readout. The physical observation for an MCP interacting with the detector is a very small amount of ionization and scintillation light. Based on Ref.~\cite{Patrignani:2016xqp}, a one-electric-charge minimum-ionizing-particle would generate roughly $10^{6}$ photoelectrons (PE) in a 1 m plastic scintillator with a density of $\rho_\mathrm{scint.} = 1$ g/cm$^3$. As a rough estimate, the number of PE produced in the scintillator scales with $\varepsilon^2$, so we can expect sensitivity to reach $\varepsilon \sim 10^{-3}$ if single-PE readout is achievable. Each scintillator stack will consist of 400 \mbox{5 cm} $\times$ 5 cm $\times$ 80 cm scintillator bars coupled to PMTs. Each bar will be oriented in the direction of the neutrino beam. A triple coincidence in all three adjacent scintillator bar-PMT sets in a small time window (nominally 15 nanoseconds) is required as an experimental signature to reduce the background. 

We consider a runtime corresponding to one year ($3\times 10^7$ s) of live-trigger time, correspond to either $6\times 10^{20}$ protons on target for the NuMI beam~\cite{Adamson:2015dkw} or $10^{21}$ POT for the DUNE beam~\cite{DUNECollaboration2015}. This live-trigger time is not the same as the actual runtime, which would be longer, and remains to be determined in a more detailed study. However, a one-year live trigger time of FerMINI should be easily obtained, given the small dead/live ratio estimated by milliQan~\cite{Ball:2016zrp} and the fact that NuMI will run for roughly five more years and LBNF is expected to run for more than ten years.
This time also corresponds to the proposal using the High-Luminosity LHC (HL-LHC) upgrade for milliQan and will allow us to directly compare with that sensitivity reach.

The advantage of having three separate scintillator array stacks is that, by requiring the observation of PEs in each stack in a narrow time window, we can drastically reduce the number of most background events \cite{Haas:2014dda,Ball:2016zrp}. Of course, this raises the requirement for signal detection too. We quantify this effect below. If we define $\overline{N}_\mathrm{PE}$ as the average number of photoelectrons collected due to a single minicharged particle $\chi$ passing through one stack of scintillators, then the probability of observing a single particle producing at least one PE in each stack, using a Poisson distribution, is
\begin{equation}
P = \left( 1 - e^{-\overline{N}_\mathrm{PE}}\right)^3.
\end{equation}
We can roughly estimate $\overline{N}_\mathrm{PE}$ as follows: $\overline{N}_\mathrm{PE} \sim \rho_\mathrm{scint.}L_\mathrm{scint.} \left<- \frac{dE}{dx}\right> \times y  \times E_\mathrm{det.}$, where $\rho_\mathrm{scint.}$ and $L_\mathrm{scint.}$ are the scintillator density and the single-scintillator-bar length, respectively, $\left<-\frac{dE}{dx}\right>$ is the mean rate of energy deposition (or energy loss) by a charged particle in the material\footnote{
This quantity is called the ``mass stopping power''  and the unit is in $\rm MeV g^{-1}cm^2$ in Ref.~\cite{Tanabashi:2018oca}. The ``linear stopping power'' is $\left<- \frac{dE}{dx}\right> \rho$ in the unit of MeV/cm where $\rho$ is the density in $\rm g/cm^3$.}, 
$y$ is the photon yield (number of photons per keV of energy deposited) of the scintillator, and $E_\mathrm{det.}$ is the detection efficiency (the fraction of produced photons that are measured by the PMTs)~\cite{Tanabashi:2018oca}. The mean energy deposition rate $\left<-\frac{dE}{dx}\right>$ is proportional to $\varepsilon^2$. Strictly speaking, $\overline{N}_\mathrm{PE}$ is not exactly linearly proportional to $L_\mathrm{scint.}$ and 
$\left<-\frac{dE}{dx}\right>$. First, one needs to take into account the scintillator's attenuation length (roughly 2 meters for the plastic scintillator considered here) and we also expect the detection efficiency $E_\mathrm{det.}$ to be lower as the scintillator length extends. Second, the dependence of $\overline{N}_\mathrm{PE}$ on $\left<-\frac{dE}{dx}\right>$ should also be modified according to Birks' law \cite{Birks:1951boa}. This rough estimation however captures the essence of the scintillation signature concisely.

We assume the scintillator is a Saint-Gobain BC-408 plastic scintillator and the PMTs are Hamamatsu R329-02 PMTs, as in Ref.~\cite{Ball:2016zrp} in performing calculations of these quantities, but specific scintillator and PMT choices can be later optimized for actual installation in this context. We can define a new quantity $\xi$ by $\overline{N}_\mathrm{PE} = (\varepsilon/\xi)^2$ ($\xi$ denotes the value of $\varepsilon$ for which one expects an average of one photoelectron per detector stack per MCP) and find $\xi \sim 2\times 10^{-3}$ for the $m_\chi$ range of interest in Ref. \cite{Ball:2016zrp}. This agrees with the explorations in the literature and more detailed GEANT simulations therein. In producing Figs.~\ref{fig:MoneyPlot_NuMI} and \ref{fig:MoneyPlot_DUNE}, we allow $\overline{N}_\mathrm{PE}$ to vary depending on the mass and energy of the $\chi$ being produced. The standard designs of FerMINI at MINOS and FerMINI at DUNE yield almost the same sensitivity, however it is by accident; this is because the DUNE setup allows for a larger number of protons on target and a closer detector, where at MINOS, there is the secondary production of MCP at the absorber closer to the detector.

Given the much larger flux of MCP produced in a proton fixed-target facility compared to that by the LHC for MCP below $\sim$ 10 GeV \cite{Magill:2018tbb}, one can potentially probe MCP with fractional charges one-order below $\varepsilon \sim 10^{-3}$. This advantage, in turn, brings up two complications: first, one should consider the most probable energy loss instead of the average energy loss in this case~\cite{Tanabashi:2018oca}. Secondly, one would be sensitive to small enough values of $\varepsilon$ that the expected number of photons produced per MCP passing through the detector is less than one, pushing the limit of the scintillator itself.

Let's consider the first complication.
Similar to considering a small-thickness detector, one should indeed consider ``the most probable energy loss" \cite{Tanabashi:2018oca} in stead of averaged energy loss described by Bethe-Bloch equation in the situation of very sparse collisions~\footnote{The scattering probability scales as $\varepsilon^2$, so small $\varepsilon$ implies sparse collisions.}. This is because the hard collisions (with large energy transfer) of a single MCP become very rare in finite length scintillator due to the very small charge, the energy loss probability distribution is highly-skewed towards the lower energy end (normally, the energy loss probability distribution is much less skewed for a $\sim$1 meter plastic scintillator since one usually does not consider such minicharged carriers).
This leads to smaller values of $\left<-\frac{dE}{dx}\right>$ when $\varepsilon \lesssim 10^{-2}$. We find that $\overline{N}_\mathrm{PE}$ is lower by a factor of roughly $2$ when $\varepsilon \simeq 10^{-4}$ (compared to the na{\"i}ve expectation), and incorporate this effect into our calculation when producing Figs.~\ref{fig:MoneyPlot_NuMI} and \ref{fig:MoneyPlot_DUNE}.

Secondly, for $\varepsilon$ much smaller than $10^{-3}$, we potentially reach the detection limitation of the nominal scintillator. As discussed in Refs.~\cite{Haas:2014dda,Ball:2016zrp}, a single MCP with very small charge would produce less than one photon in the scintillator on average. In principle, this just means the chance of triple coincidence of the MCP in the scintillator is very small according to the Poisson distribution. However, the scintillator performance for such small numbers of produced photons has not been fully explored. We thus consider the expected number of photons generated by one MCP in a scintillator bar,
$\overline{N}_{\mathrm{s}\gamma} (\varepsilon) \equiv \overline{N}_\mathrm{PE}/E_\mathrm{det.}$.
We denote the region for $\varepsilon$ below which $\overline{N}_{\mathrm{s}\gamma} \lesssim 1$ in Figs.~\ref{fig:MoneyPlot_NuMI} and \ref{fig:MoneyPlot_DUNE} by dashed curves. One can take the regime above these curves to be a conservative estimate of FerMINI sensitivity -- even if we restrict ourselves to $\overline{N}_{\mathrm{s}\gamma} \gtrsim 1$, this sensitivity is still very competitive in the range $10$ MeV $\lesssim m_\chi \lesssim 5$ GeV. A more thorough investigation should allow for analysis including the $\overline{N}_{\mathrm{s}\gamma} < 1$ region~\cite{MCP_Proposal}.

Finally, let us consider the possibilities to increase the number of photoelectrons produced in the detector, and thus probe even smaller values of $\varepsilon$ given the advantage of high flux of MCPs produced in the proton beam.
One can either elongate the scintillator bars, use an alternative material that has higher light yield, or dope the scintillator bar to enhance the light yield (see the development of ``Wunderbar'' technology~\cite{wunderbar1,wunderbar2}.) --
in fact, the milliQan collaboration has considered constructing the detector with other materials to increase $\overline{N}_\mathrm{PE}$, including sodium iodide (NaI), lanthanum bromide (LaBr$_3$), or liquid xenon (Xe). These materials all have higher photon yield per energy $y$ but also higher costs~\cite{milliQan:slides}.
We look into the sensitivity gain by enhancing the overall scintillating capability and thus increase the $\overline{N}_{PE}$ by five-fold in the FerMINI setup.
For FerMINI at MINOS hall, we consider such possibility of enhancing the $\overline{N}_{PE}$ by five times but reduce the number of scintillator bar-PMT sets by one-fifth to na{\"i}vely balance the cost. This roughly reduces the number of MCP traversing the detector by a factor of five, but the increase in $\overline{N}_{PE}$  more than compensates for the lower flux.
This modified detector's sensitivity is shown in comparison with the nominal design in Fig.~\ref{fig:MoneyPlot_NuMI}.
Additionally, we explore the idea of simply increasing the $\overline{N}_{PE}$ by five times and keeping the cross-sectional area constant for FerMINI at the DUNE near detector hall. This is shown in comparison to the nominal design in Fig.~\ref{fig:MoneyPlot_DUNE}. We find that the modified detector at MINOS increases sensitivity to $\varepsilon$ by roughly a factor of two for $10$ MeV $\lesssim m_\chi \lesssim 1$ GeV, where the enhanced detector at DUNE improves sensitivity, unsurprisingly, for all MCP masses. 
Note that the choice of five times greater scintillating capacity could be very optimistic and it is mainly a judicial choice of demonstrating the improvement of sensitivity in Fig.~\ref{fig:MoneyPlot}.

Note that, even though we attempt to take into account the subtleties in producing the sensitivity curves, a detailed simulation is needed to have a realistic estimation of the FerMINI projection. Finally, while one could also worry about the attenuation and angular deflection of MCP produced in the target that traverse dirt, absorber, and rock en route to the detector, we find such effects to be negligible here. For $10$ MeV $\lesssim m_\chi \lesssim 5$ GeV, the average $\chi$ energy is several GeV, meaning that the total energy loss due to such attenuation is insignificant. The attenuation of MCP flux has been explored in detail in Ref.~\cite{Prinz:2001qz} and has been shown to be small for $m_\chi \gtrsim 1$ MeV as long as $\varepsilon < 10^{-1}$. 
The angular deflection of $\chi$ is estimated in Refs.~\cite{Precite_Roni,Roni:slides} and is insignificant for FerMINI (where we require that the total deflection is less than $\Delta \theta \lesssim \frac{5\ \mathrm{cm}}{80\ \mathrm{cm}}$) as long as $\varepsilon < 10^{-1}$.
We will conduct a detailed study including detector simulations and passage through the materials for specific experimental sites in Ref.~\cite{MCP_Proposal}.

\section{Background $\&$ Sensitivity}
\label{sec:Sensitivity}

In this section, we discuss the background of the FerMINI setup and the sensitivity projection. We will separate the discussions of the background roughly into two categories: detector background and beam-related background. The detector background includes the sources that are present for the detector without the existence of the beam. The beam related background is the additional background that may be generated once the proton beam is turned on.

First, we look at the detector background. Since our nominal detector design is the same as that of the milliQan \cite{Haas:2014dda,Ball:2016zrp}, we share the same detector background and can use the strategy they developed to suppress such backgrounds. Three major backgrounds were pointed out in the milliQan literature: dark currents in the PMT (most dominant), natural radiation background (e.g. cosmic muons), and PMT after-pulses. Four major strategies were developed \cite{Haas:2014dda,Ball:2016zrp} to suppress the background as summarized below:

\begin{itemize}
\setlength\itemsep{-0.25em}
\item Require triple coincidence as a detection signature:\\ 
Requiring at least one PE in each PMT connected to the three contiguous scintillators in each of the three stacks of scintillators within a $15$ ns window. This would help reduce all background sources except those caused by SM charged particles.
\item Offline-vetoes of large-PE ($>$ 10 PE) events:\\ Radiation background (e.g., cosmic muons) traversing the detector produces a large number of PE (typically more than $10^3$). An offline veto of events with more than 10 PEs would veto these events. 
\item Offset the middle detector array:\\ Charged SM particles (again, such as muons) could skim through the edge of the detector, producing low numbers of PEs, and thus be confused as signatures. Shifting the middle detector, or making it slightly larger (with $\mathcal{O}(10)$ more scintillator-PMT sets), would help to avoid such particle trajectories.
\item Deadtime veto of the after-pulses:\\ When a large pulse ( corresponding to more than 10 PEs) enters the PMT, a smaller after-pulse can be created that may survive the aforementioned large-PE veto. However, since these after-pulses are generated within roughly 10 microseconds, it will stay within the deadtime of the readout board (considered in Ref.~\cite{Ball:2016zrp}) triggered by the large pulse and thus not being recorded.
\end{itemize}

After these background reductions, the detector background would be greatly reduced. The remaining dominating background would be the triple coincidence dark current counts in the PMT. Assuming a dark current rate to be roughly $\nu$ = 500 Hz in the choice of PMTs~\cite{Ball:2016zrp} (which can be further reduced to 80 Hz by cooling the system to $-20^\circ$ C~\cite{milliQan:slides}), the corresponding background rate for each triple PMT set is $\nu_B= \nu^3 (\Delta t)^2 = 2.8 \times 10^{-8}$ Hz, for the time window $\Delta t = 15$ ns. With 400 PMT sets, this gives a background rate of roughly 300 events per year\footnote{In Section~\ref{sec:Setup} we discussed the possibility of increasing the number of photoelectrons $\overline{N}_\mathrm{PE}$ by improving the scintillator and decreasing the number of scintillator bar-PMT sets. In this alternate design, the number of background events per year from dark current would also drop roughly by a factor of five, given that there are five times fewer PMT sets.}. Additional strategies can be applied to further reduce the background as discussed in Ref.~\cite{Ball:2016zrp}, but we directly take 300 events for the detector background of the nominal design of FerMINI as a rather conservative estimation of the background for the sensitivity projection.

Let us move on to the beam-related background.
Here we briefly discuss backgrounds that are related to the intense proton beams: SM charged particles that reach, or are being produced in the detector, and neutral particles like neutrons and neutrinos.
SM charged particles can be produced directly from the beam on target, through secondary production in the dirt and rock, or through neutral particles (e.g., neutron and neutrino) and charged particles scattering in the detector, and potentially generate the triple-coincidence events.
The SM charged-particle flux produced at target or through secondary productions would be shielded by the layers of rocks and monitored by the muon monitors before reaching the detector.  
The charged particles that arrive at or are produced in the detector would then be vetoed the same techniques as discussed above, by vetoing large-PE events and offset the middle detector. 
Electrons and protons (or other charged particles, e.g. pions) from hard neutrino and neutron scattering would also be vetoed as large-PE events, but the soft collisions (e.g. neutrino neutral-current scattering) could mimic the low-PE ionization signature.
As for these neutrino-related events, in a plastic detector of total volume $1$ m $\times 1$ m $\times $ 3 m, we expect $\mathcal{O}(10^4)$ neutral-current quasi-elastic scattering events per year~\cite{DUNECollaboration2015}. 
A rough estimate shows that requiring soft events in each detector array in a 15 ns window gives an expected number of $\mathcal{O}(10^{-18})$ background events for one year of live-trigger time, meaning this background is negligible in either the NuMI or DUNE beamlines. 
For the sensitivity curves in this plot, we take the beam-related background to be of the same order as the detector background, i.e., 300 events for the nominal 1 m $\times$ 1 m $\times$ 3 m configuration. For the modified designs at MINOS and at DUNE, we scale up the beam-related background according to the modified overall scintillating capacities of the detectors as rough estimates.
 
A realistic determination of the beam-related background at Fermilab beams would require a detailed simulation including the beam production, the configuration between the target and the detector, and the detector itself. All in all, the flux of the SM charged particles should be able to be determined with the help of the NuMI-MINOS and LBNF-DUNE collaborations. An additional way to monitor beam-related backgrounds would be to utilize the movable DUNE PRISM detector proposal~\cite{duneprism}. Because the MCP come (predominantly) from neutral meson decay, where the beam-related backgrounds come from charged meson decay, the focusing magnets will generate a narrow neutrino/charged particle beam, where the MCP flux would be broader. Making measurements at different off-axis angles would allow one to disentangle the MCP signal from beam-related backgrounds. Lastly, a beam-dump type experiment using the NuMI or DUNE beam (similar to that performed with the Booster beam for MiniBooNE~\cite{Aguilar-Arevalo:2018wea}) would aide in such a determination. We leave these possibilities to a more detailed study~\cite{MCP_Proposal}.

As one additional note regarding backgrounds, \mbox{FerMINI} at the NuMI/MINOS site (roughly 100 m underground), FerMINI at the DUNE site (roughly 60 m underground), and milliQan at the CERN site (roughly 70 m underground) all have reduced cosmic background rates, but the shieldings are not ideal comparing to that of the deep underground labs (e.g., Laboratori Nazionali del Gran Sasso \cite{Miramonti:2005xq}).
We plan to conduct in-situ background measurement at Fermilab sites with a scaled-down prototype detector to better estimate rates of each background events, and rely on the aforementioned techniques (especially the large-PE veto and offsetting middle detector) to suppress these background.


\begin{figure*}[!htbp]
\begin{subfigure}[!htbp]{0.485\textwidth}
    \includegraphics[width=\textwidth]{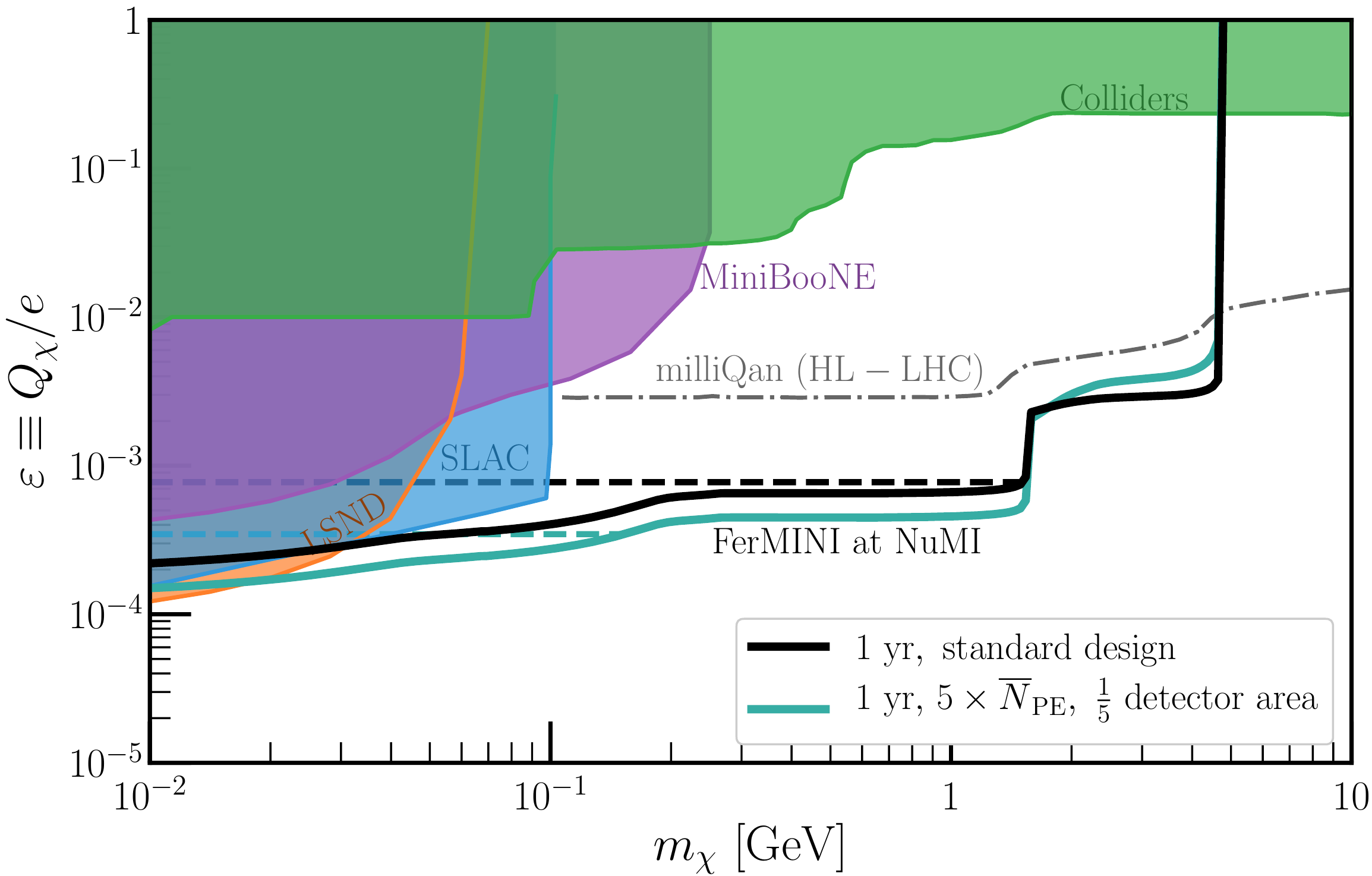}
    \caption{The sensitivity of FerMINI in the NuMI Beamline at the MINOS Near Detector site.}
    \label{fig:MoneyPlot_NuMI}
  \end{subfigure}
  \begin{subfigure}[!htbp]{0.485\textwidth}
    \includegraphics[width=\textwidth]{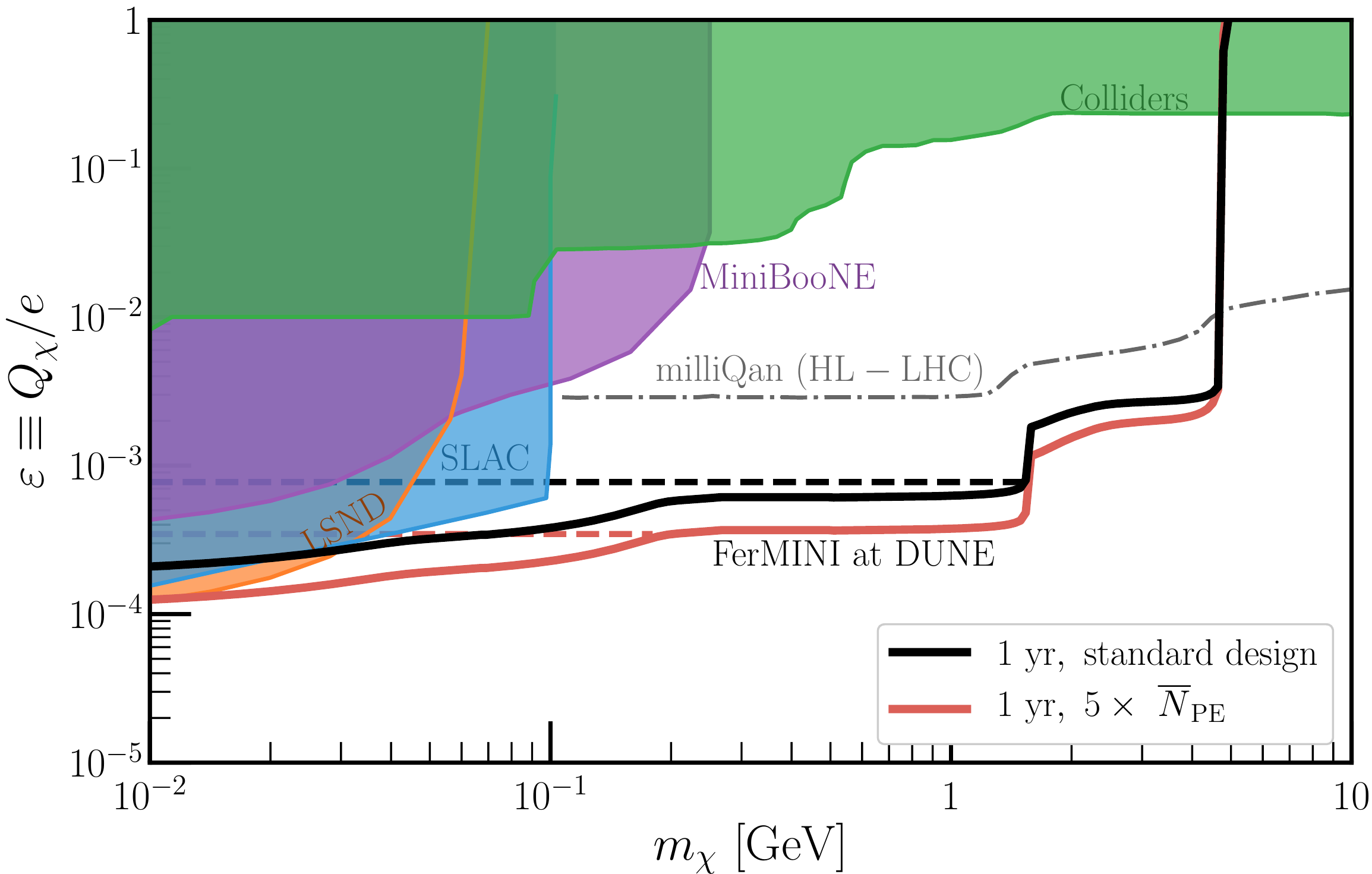}
    \caption{The sensitivity of FerMINI in the DUNE Near Detector hall.}
    \label{fig:MoneyPlot_DUNE}
  \end{subfigure}
  \caption{Expected 95\% CL sensitivity to minicharged particles in the two potential sites we consider. The solid curves are the sensitivity reaches of FerMINI at each site operating for one year. The black, solid curves are based on the nominal design in both sites. The light blue curve in (a) is the sensitivity of a detector with alternative design (see Section \ref{sec:Setup} for detail) with one year of operation in the MINOS near detector hall. The light red curve in (b) is the projection of a detector with 5 times better scintillation capability and unchanged detection area placed in the DUNE near detector hall, again operating for one year. The milliQan HL-LHC sensitivity reach \cite{Haas:2014dda,Ball:2016zrp} is plotted as dot-dashed curve for comparison. The shaded regimes are the existing constraints from SLAC \cite{Prinz:1998ua}, collider \cite{Davidson:2000hf}, MiniBooNE and LSND \cite{Magill:2018tbb}. For each FerMINI sensitivity curve, we include a dashed curve indicating values of $\varepsilon$ corresponding to $\overline{N}_{\mathrm{s}\gamma} \approx 1$ (see Section \ref{sec:Setup} for detail). The sensitivity reaches below these dashed lines require more detailed analyses. The reason why standard designs of FerMINI at NuMI and at DUNE yield very similar sensitivity is explained in text  (see also discussions of NuMI secondary beam productions at the absorbers in Section \ref{sec:Setup}).
  }\label{fig:MoneyPlot} 
\end{figure*}

In Figs.~\ref{fig:MoneyPlot_NuMI} and \ref{fig:MoneyPlot_DUNE} we show 95\% CL sensitivity following sensitivity analysis in \cite{Tanabashi:2018oca} assuming aforementioned background events and 1 year of live-trigger time at NuMI and DUNE, respectively. We also show, as discussed above, the alternative FerMINI designs that lead to improved sensitivity. We plan to apply for support from the Fermilab Laboratory Directed Research and Development (LDRD) Program, allowing for construction and development of a prototype detector for in situ measurements on-site at Fermilab. A prototype detector, even if constructed in the next few years, also has the possibility of constraining parameter space thanks to the high flux at NuMI and its consistent running in the coming years for the variety of neutrino experiments it provides for.

Finally, given the large flux of MCP at neutrino beams, one can consider adding an additional stack of 400 scintillator-PMTs and require quadruple coincidence as an experimental signature to further cut down the background. This would inevitably sacrifice event rates but theoretically/naively reduce the dark current background by a factor of $\nu \Delta t \sim 10^{-5}$, in turn reducing the dark current background to zero events per year. We have explored the potential of such a setup and find roughly the same sensitivity reach as with three detector stacks; however, this would be an experimental demonstration or test of a zero dark-current background search.

\section{New Detector Design with Neutrino Detectors}
\label{sec:Alternatives}

We would like to discuss a few alternative options beyond the nominal FerMINI proposal that could potentially provide comparable or even better sensitivity and mass reach for MCP.

Since the detector will be located inside existing/future neutrino near detector halls, we discuss the possibility of utilizing the neutrino detectors to better enhance our sensitivity to MCP. One idea would be to place FerMINI directly in front of or behind the neutrino detector and using the liquid argon technology to provide extra information on particles traversing the MCP detector and provide a veto on SM particles that could fake an MCP signal. In addition, one can look for mixed signature combining the scintillation signature discussed in this work and the hard electron scattering signature utilizing liquid argon or Cherenkov neutrino detectors discussed in \cite{Magill:2018tbb}.

In particular, the excellent resolution of liquid argon could also be leveraged, by splitting the FerMINI detector arrays, placing some in front of and some behind the neutrino detector. Minicharged particles traveling through liquid argon can scatter off electrons, leading to single-electron events~\cite{Magill:2018tbb,Precite_Roni}. The combination of this signature with the scintillation signal of FerMINI could potentially further improve the MCP sensitivity. Both the NuMI and DUNE locations could provide such a combination, using the existing ArgoNeuT detector \cite{Acciarri:2018myr} and the upcoming DUNE near detector. 
In addition, the DUNE near detector complex plans to have a 3D Scintillator Tracker (3DST) that could also be leveraged along with FerMINI to look for more distinctive signatures of MCP and other new physics scenarios.
These experimental designs are left for further exploration~\cite{MCP_Proposal}.

An alternative site to host a dedicated MCP detector is the CERN Super Proton Synchroton (SPS) facility. Specifically, the SPS has 400 GeV of beam energy (compared to $120$ GeV at NuMI/DUNE), allowing us to search for heavier MCP\footnote{The maximum $m_\chi$ probed in a proton fixed-target experiment is set by $m_\chi^\mathrm{(max.)} = \sqrt{s}/2 = \sqrt{2 m_p E_\mathrm{beam}}/2$. For $E_\mathrm{beam} = 120$ ($400$) GeV, this gives $m_\chi^\mathrm{(max.)} \simeq 7.5$ ($13.7$) GeV.}~\cite{Magill:2018tbb}. The number of $\chi$ produced via Drell-Yan production (See the gray curve in Fig.~\ref{fig:ChiProduction}) would be enhanced in this scenario. 
There is potentially room to host an MCP detector in the structures hosting various experiments, including NA62 \cite{NA62:2017rwk} and the proposed Search for Hidden Particles (SHiP) \cite{SHiP:2018yqc} experiments. The details including shielding and space would have to be studied in each potential sites.
One can again combine the MCP detector with the existing and proposed experiments, especially the SHiP experiment to get better signature recognition or background suppression.

\section{Discussion $\&$ Conclusion}

In this section, we briefly discuss the advantages of FerMINI and future prospects.

Compared to milliQan, our setup consists of a much larger flux of MCP reaching the detector, due to the higher proton beam intensity.
We find that FerMINI is sensitive to $\varepsilon$ below $10^{-3}$ and has better sensitivity than the milliQan search with the HL-LHC up to about $m_\chi \sim 5$ GeV. The MCP flux is so large that it potentially saturates the scintillation limit and force us to consider modified detectors to fully explore the potential of FerMINI.

Another great advantage of FerMINI is that the NuMI beam is currently under operation. 
The LHC is under a long shutdown and milliQan has to wait for the LHC upgrades to complete in roughly two years to resume operation. The NuMI beam will be shut down in roughly five years, so it is important to initiate the FerMINI construction as soon as possible to take full advantage of the NuMI beam operation.

In addition, with a small detector and simple technology originally developed by milliQan, the FerMINI proposal makes for a great movable addition to current and upcoming Fermilab experiments, specifically the DUNE experiment. 
One can imagine developing the detector in one site and moving it for a longer-term operation. This is specifically advantageous at Fermilab, where it can be used in the NuMI beam until its shutdown in roughly five years, then moved to the DUNE near detector hall.

The milliQan Collaboration has invested in and achieved a great understanding of the detector performance and have constructed a $1 \%$ demonstrator to conduct a test run at CERN \cite{Yoo:2018lhk}. As mentioned above, since the LHC is entering a long shutdown in 2018, it is natural to explore the possibility of moving this prototype to Fermilab or the CERN SPS to conduct in-situ measurements, especially to aide in background estimation, but also to potentially search for MCP. However, since this detector technology is low-cost, it may be favorable to directly construct a new prototype at Fermilab. 
As mentioned above, we plan to apply for support from the Fermilab Laboratory Directed Research and Development (LDRD) Program, allowing for construction and development of a prototype detector for in situ measurements on-site at Fermilab.
Such a prototype development could also allow for exploring the alternatives discussed in Section~\ref{sec:Sensitivity}.
We regard the implementations of this proposal as a great opportunity for synergy between the collider, neutrino, and dark matter communities at Fermilab and at CERN.

\section{Acknowledgments}
We thank Joachim Kopp, Gabriel Magill, Zarko Pavlovic, Ryan Plestid, and Maxim Pospelov for useful discussions and input on various stages of completing this work. We especially thank Maxim Pospelov for the discussions of FerMINI alternative designs.
We also thank members in the milliQan collaboration, specifically Albert De Roeck, Andy Haas, Christopher Hill, and Max Swiatlowski, for direct and indirect correspondence regarding the details of the milliQan proposal.
We also thank Roni Harnik for pointing out the subtlety in determining the energy loss of a charged particle with very rare collisions in the detector, and Patrick Fox for correspondence regarding the NuMI beamline. We also thank Carlos Arg{\"u}elles, Jeffrey Berryman, Kareem Farrag, Noah Kurinsky, Shirley Li, Mark Ross-Lonergan, Ibrahim Safa, and Gary Shiu for useful references and feedback on our draft.
YT thanks CERN, where this work was conceived, and its theory group for the hospitality.

This manuscript has been authored by Fermi Research Alliance, LLC under Contract No. DE-AC02-07CH11359 with the U.S. Department of Energy, Office of Science, Office of High Energy Physics.

\bibliography{Numi_MCP}

\begin{thebibliography}{66}
\expandafter\ifx\csname natexlab\endcsname\relax\def\natexlab#1{#1}\fi
\expandafter\ifx\csname bibnamefont\endcsname\relax
  \def\bibnamefont#1{#1}\fi
\expandafter\ifx\csname bibfnamefont\endcsname\relax
  \def\bibfnamefont#1{#1}\fi
\expandafter\ifx\csname citenamefont\endcsname\relax
  \def\citenamefont#1{#1}\fi
\expandafter\ifx\csname url\endcsname\relax
  \def\url#1{\texttt{#1}}\fi
\expandafter\ifx\csname urlprefix\endcsname\relax\def\urlprefix{URL }\fi
\providecommand{\bibinfo}[2]{#2}
\providecommand{\eprint}[2][]{\url{#2}}

\bibitem[{\citenamefont{Dirac}(1931)}]{Dirac:1931kp}
\bibinfo{author}{\bibfnamefont{P.~A.~M.} \bibnamefont{Dirac}}, ``{Quantized
  Singularities in the Electromagnetic Field},'' \bibinfo{journal}{Proc. Roy.
  Soc. Lond.} \textbf{\bibinfo{volume}{A133}}, \bibinfo{pages}{60}
  (\bibinfo{year}{1931}), \bibinfo{note}{[,278(1931)]}.

\bibitem[{\citenamefont{Pati and Salam}(1973)}]{Pati:1973uk}
\bibinfo{author}{\bibfnamefont{J.~C.} \bibnamefont{Pati}} \bibnamefont{and}
  \bibinfo{author}{\bibfnamefont{A.}~\bibnamefont{Salam}}, ``{Unified
  Lepton-Hadron Symmetry and a Gauge Theory of the Basic Interactions},''
  \bibinfo{journal}{Phys. Rev.} \textbf{\bibinfo{volume}{D8}},
  \bibinfo{pages}{1240} (\bibinfo{year}{1973}).

\bibitem[{\citenamefont{Georgi and Glashow}(1974)}]{Georgi:1974sy}
\bibinfo{author}{\bibfnamefont{H.}~\bibnamefont{Georgi}} \bibnamefont{and}
  \bibinfo{author}{\bibfnamefont{S.~L.} \bibnamefont{Glashow}}, ``{Unity of All
  Elementary Particle Forces},'' \bibinfo{journal}{Phys. Rev. Lett.}
  \textbf{\bibinfo{volume}{32}}, \bibinfo{pages}{438} (\bibinfo{year}{1974}).

\bibitem[{\citenamefont{Dobroliubov and Ignatiev}(1990)}]{Dobroliubov:1989mr}
\bibinfo{author}{\bibfnamefont{M.~I.} \bibnamefont{Dobroliubov}}
  \bibnamefont{and} \bibinfo{author}{\bibfnamefont{A.~{\relax Yu}.}
  \bibnamefont{Ignatiev}}, ``{MILLICHARGED PARTICLES},''
  \bibinfo{journal}{Phys. Rev. Lett.} \textbf{\bibinfo{volume}{65}},
  \bibinfo{pages}{679} (\bibinfo{year}{1990}).

\bibitem[{\citenamefont{Prinz et~al.}(1998)}]{Prinz:1998ua}
\bibinfo{author}{\bibfnamefont{A.~A.} \bibnamefont{Prinz}}
  \bibnamefont{et~al.}, ``{Search for millicharged particles at SLAC},''
  \bibinfo{journal}{Phys. Rev. Lett.} \textbf{\bibinfo{volume}{81}},
  \bibinfo{pages}{1175} (\bibinfo{year}{1998}), \eprint{hep-ex/9804008}.

\bibitem[{\citenamefont{Davidson et~al.}(2000)\citenamefont{Davidson,
  Hannestad, and Raffelt}}]{Davidson:2000hf}
\bibinfo{author}{\bibfnamefont{S.}~\bibnamefont{Davidson}},
  \bibinfo{author}{\bibfnamefont{S.}~\bibnamefont{Hannestad}},
  \bibnamefont{and} \bibinfo{author}{\bibfnamefont{G.}~\bibnamefont{Raffelt}},
  ``{Updated bounds on millicharged particles},'' \bibinfo{journal}{JHEP}
  \textbf{\bibinfo{volume}{05}}, \bibinfo{pages}{003} (\bibinfo{year}{2000}),
  \eprint{hep-ph/0001179}.

\bibitem[{\citenamefont{Prinz}(2001)}]{Prinz:2001qz}
\bibinfo{author}{\bibfnamefont{A.~A.} \bibnamefont{Prinz}}, Ph.D. thesis,
  \bibinfo{school}{Stanford U., Phys. Dept.} (\bibinfo{year}{2001}),
  \urlprefix\url{http://wwwlib.umi.com/dissertations/fullcit?p3002033}.

\bibitem[{\citenamefont{Golowich and Robinett}(1987)}]{Golowich1987}
\bibinfo{author}{\bibfnamefont{E.}~\bibnamefont{Golowich}} \bibnamefont{and}
  \bibinfo{author}{\bibfnamefont{R.~W.} \bibnamefont{Robinett}}, ``{Limits on
  Millicharged Matter From Beam Dump Experiments},'' \bibinfo{journal}{Phys.
  Rev.} \textbf{\bibinfo{volume}{D35}}, \bibinfo{pages}{391}
  (\bibinfo{year}{1987}).

\bibitem[{\citenamefont{Babu et~al.}(1994)\citenamefont{Babu, Gould, and
  Rothstein}}]{Babu:1993yh}
\bibinfo{author}{\bibfnamefont{K.~S.} \bibnamefont{Babu}},
  \bibinfo{author}{\bibfnamefont{T.~M.} \bibnamefont{Gould}}, \bibnamefont{and}
  \bibinfo{author}{\bibfnamefont{I.~Z.} \bibnamefont{Rothstein}}, ``{Closing
  the windows on MeV Tau neutrinos},'' \bibinfo{journal}{Phys. Lett.}
  \textbf{\bibinfo{volume}{B321}}, \bibinfo{pages}{140} (\bibinfo{year}{1994}),
  \eprint{hep-ph/9310349}.

\bibitem[{\citenamefont{Gninenko et~al.}(2007)\citenamefont{Gninenko,
  Krasnikov, and Rubbia}}]{Gninenko:2006fi}
\bibinfo{author}{\bibfnamefont{S.~N.} \bibnamefont{Gninenko}},
  \bibinfo{author}{\bibfnamefont{N.~V.} \bibnamefont{Krasnikov}},
  \bibnamefont{and} \bibinfo{author}{\bibfnamefont{A.}~\bibnamefont{Rubbia}},
  ``{Search for millicharged particles in reactor neutrino experiments: A Probe
  of the PVLAS anomaly},'' \bibinfo{journal}{Phys. Rev.}
  \textbf{\bibinfo{volume}{D75}}, \bibinfo{pages}{075014}
  (\bibinfo{year}{2007}), \eprint{hep-ph/0612203}.

\bibitem[{\citenamefont{Chatrchyan et~al.}(2013)}]{CMS:2012xi}
\bibinfo{author}{\bibfnamefont{S.}~\bibnamefont{Chatrchyan}}
  \bibnamefont{et~al.} (\bibinfo{collaboration}{CMS}), ``{Search for
  fractionally charged particles in $pp$ collisions at $\sqrt{s}=7$ TeV},''
  \bibinfo{journal}{Phys. Rev.} \textbf{\bibinfo{volume}{D87}},
  \bibinfo{pages}{092008} (\bibinfo{year}{2013}), \eprint{1210.2311}.

\bibitem[{\citenamefont{Agnese et~al.}(2015)}]{Agnese:2014vxh}
\bibinfo{author}{\bibfnamefont{R.}~\bibnamefont{Agnese}} \bibnamefont{et~al.}
  (\bibinfo{collaboration}{CDMS}), ``{First Direct Limits on Lightly Ionizing
  Particles with Electric Charge Less Than $e/6$},'' \bibinfo{journal}{Phys.
  Rev. Lett.} \textbf{\bibinfo{volume}{114}}, \bibinfo{pages}{111302}
  (\bibinfo{year}{2015}), \eprint{1409.3270}.

\bibitem[{\citenamefont{Haas et~al.}(2015)\citenamefont{Haas, Hill, Izaguirre,
  and Yavin}}]{Haas:2014dda}
\bibinfo{author}{\bibfnamefont{A.}~\bibnamefont{Haas}},
  \bibinfo{author}{\bibfnamefont{C.~S.} \bibnamefont{Hill}},
  \bibinfo{author}{\bibfnamefont{E.}~\bibnamefont{Izaguirre}},
  \bibnamefont{and} \bibinfo{author}{\bibfnamefont{I.}~\bibnamefont{Yavin}},
  ``{Looking for milli-charged particles with a new experiment at the LHC},''
  \bibinfo{journal}{Phys. Lett.} \textbf{\bibinfo{volume}{B746}},
  \bibinfo{pages}{117} (\bibinfo{year}{2015}), \eprint{1410.6816}.

\bibitem[{\citenamefont{Ball et~al.}(2016)}]{Ball:2016zrp}
\bibinfo{author}{\bibfnamefont{A.}~\bibnamefont{Ball}} \bibnamefont{et~al.},
  ``{A Letter of Intent to Install a milli-charged Particle Detector at LHC
  P5},''  (\bibinfo{year}{2016}), \eprint{1607.04669}.

\bibitem[{\citenamefont{Alvis et~al.}(2018)}]{Alvis:2018yte}
\bibinfo{author}{\bibfnamefont{S.~I.} \bibnamefont{Alvis}} \bibnamefont{et~al.}
  (\bibinfo{collaboration}{Majorana}), ``{First Limit on the Direct Detection
  of Lightly Ionizing Particles for Electric Charge as Low as e/1000 with the
  Majorana Demonstrator},'' \bibinfo{journal}{Phys. Rev. Lett.}
  \textbf{\bibinfo{volume}{120}}, \bibinfo{pages}{211804}
  (\bibinfo{year}{2018}), \eprint{1801.10145}.

\bibitem[{\citenamefont{Magill et~al.}(2018{\natexlab{a}})\citenamefont{Magill,
  Plestid, Pospelov, and Tsai}}]{Magill:2018tbb}
\bibinfo{author}{\bibfnamefont{G.}~\bibnamefont{Magill}},
  \bibinfo{author}{\bibfnamefont{R.}~\bibnamefont{Plestid}},
  \bibinfo{author}{\bibfnamefont{M.}~\bibnamefont{Pospelov}}, \bibnamefont{and}
  \bibinfo{author}{\bibfnamefont{Y.-D.} \bibnamefont{Tsai}}, ``{Millicharged
  particles in neutrino experiments},''  (\bibinfo{year}{2018}{\natexlab{a}}),
  \eprint{1806.03310}.

\bibitem[{\citenamefont{Shiu et~al.}(2013)\citenamefont{Shiu, Soler, and
  Ye}}]{Shiu:2013wxa}
\bibinfo{author}{\bibfnamefont{G.}~\bibnamefont{Shiu}},
  \bibinfo{author}{\bibfnamefont{P.}~\bibnamefont{Soler}}, \bibnamefont{and}
  \bibinfo{author}{\bibfnamefont{F.}~\bibnamefont{Ye}}, ``{Milli-Charged Dark
  Matter in Quantum Gravity and String Theory},'' \bibinfo{journal}{Phys. Rev.
  Lett.} \textbf{\bibinfo{volume}{110}}, \bibinfo{pages}{241304}
  (\bibinfo{year}{2013}), \eprint{1302.5471}.

\bibitem[{\citenamefont{Alexander et~al.}(2016)}]{Alexander:2016aln}
\bibinfo{author}{\bibfnamefont{J.}~\bibnamefont{Alexander}}
  \bibnamefont{et~al.} (\bibinfo{year}{2016}), \eprint{1608.08632},
  \urlprefix\url{https://inspirehep.net/record/1484628/files/arXiv:1608.08632.pdf}.

\bibitem[{\citenamefont{Holdom}(1986)}]{Holdom:1985ag}
\bibinfo{author}{\bibfnamefont{B.}~\bibnamefont{Holdom}}, ``{Two U(1)'s and
  Epsilon Charge Shifts},'' \bibinfo{journal}{Phys. Lett.}
  \textbf{\bibinfo{volume}{166B}}, \bibinfo{pages}{196} (\bibinfo{year}{1986}).

\bibitem[{\citenamefont{Brahm and Hall}(1990)}]{Brahm:1989jh}
\bibinfo{author}{\bibfnamefont{D.~E.} \bibnamefont{Brahm}} \bibnamefont{and}
  \bibinfo{author}{\bibfnamefont{L.~J.} \bibnamefont{Hall}}, ``{U(1)-prime DARK
  MATTER},'' \bibinfo{journal}{Phys. Rev.} \textbf{\bibinfo{volume}{D41}},
  \bibinfo{pages}{1067} (\bibinfo{year}{1990}).

\bibitem[{\citenamefont{Boehm and Fayet}(2004)}]{Boehm:2003hm}
\bibinfo{author}{\bibfnamefont{C.}~\bibnamefont{Boehm}} \bibnamefont{and}
  \bibinfo{author}{\bibfnamefont{P.}~\bibnamefont{Fayet}}, ``{Scalar dark
  matter candidates},'' \bibinfo{journal}{Nucl. Phys.}
  \textbf{\bibinfo{volume}{B683}}, \bibinfo{pages}{219} (\bibinfo{year}{2004}),
  \eprint{hep-ph/0305261}.

\bibitem[{\citenamefont{Pospelov et~al.}(2008)\citenamefont{Pospelov, Ritz, and
  Voloshin}}]{Pospelov:2007mp}
\bibinfo{author}{\bibfnamefont{M.}~\bibnamefont{Pospelov}},
  \bibinfo{author}{\bibfnamefont{A.}~\bibnamefont{Ritz}}, \bibnamefont{and}
  \bibinfo{author}{\bibfnamefont{M.~B.} \bibnamefont{Voloshin}}, ``{Secluded
  WIMP Dark Matter},'' \bibinfo{journal}{Phys. Lett.}
  \textbf{\bibinfo{volume}{B662}}, \bibinfo{pages}{53} (\bibinfo{year}{2008}),
  \eprint{0711.4866}.

\bibitem[{\citenamefont{Bjorken et~al.}(2009)\citenamefont{Bjorken, Essig,
  Schuster, and Toro}}]{Bjorken2009}
\bibinfo{author}{\bibfnamefont{J.~D.} \bibnamefont{Bjorken}},
  \bibinfo{author}{\bibfnamefont{R.}~\bibnamefont{Essig}},
  \bibinfo{author}{\bibfnamefont{P.}~\bibnamefont{Schuster}}, \bibnamefont{and}
  \bibinfo{author}{\bibfnamefont{N.}~\bibnamefont{Toro}}, ``{New Fixed-Target
  Experiments to Search for Dark Gauge Forces},'' \bibinfo{journal}{Phys. Rev.}
  \textbf{\bibinfo{volume}{D80}}, \bibinfo{pages}{075018}
  (\bibinfo{year}{2009}), \eprint{0906.0580}.

\bibitem[{\citenamefont{Batell et~al.}(2009)\citenamefont{Batell, Pospelov, and
  Ritz}}]{Batell2009}
\bibinfo{author}{\bibfnamefont{B.}~\bibnamefont{Batell}},
  \bibinfo{author}{\bibfnamefont{M.}~\bibnamefont{Pospelov}}, \bibnamefont{and}
  \bibinfo{author}{\bibfnamefont{A.}~\bibnamefont{Ritz}}, ``{Exploring Portals
  to a Hidden Sector Through Fixed Targets},'' \bibinfo{journal}{Phys. Rev.}
  \textbf{\bibinfo{volume}{D80}}, \bibinfo{pages}{095024}
  (\bibinfo{year}{2009}), \eprint{0906.5614}.

\bibitem[{\citenamefont{deNiverville et~al.}(2011)\citenamefont{deNiverville,
  Pospelov, and Ritz}}]{deNiverville:2011it}
\bibinfo{author}{\bibfnamefont{P.}~\bibnamefont{deNiverville}},
  \bibinfo{author}{\bibfnamefont{M.}~\bibnamefont{Pospelov}}, \bibnamefont{and}
  \bibinfo{author}{\bibfnamefont{A.}~\bibnamefont{Ritz}}, ``{Observing a light
  dark matter beam with neutrino experiments},'' \bibinfo{journal}{Phys. Rev.}
  \textbf{\bibinfo{volume}{D84}}, \bibinfo{pages}{075020}
  (\bibinfo{year}{2011}), \eprint{1107.4580}.

\bibitem[{\citenamefont{Izaguirre et~al.}(2013)\citenamefont{Izaguirre,
  Krnjaic, Schuster, and Toro}}]{Izaguirre:2013uxa}
\bibinfo{author}{\bibfnamefont{E.}~\bibnamefont{Izaguirre}},
  \bibinfo{author}{\bibfnamefont{G.}~\bibnamefont{Krnjaic}},
  \bibinfo{author}{\bibfnamefont{P.}~\bibnamefont{Schuster}}, \bibnamefont{and}
  \bibinfo{author}{\bibfnamefont{N.}~\bibnamefont{Toro}}, ``{New Electron
  Beam-Dump Experiments to Search for MeV to few-GeV Dark Matter},''
  \bibinfo{journal}{Phys. Rev.} \textbf{\bibinfo{volume}{D88}},
  \bibinfo{pages}{114015} (\bibinfo{year}{2013}), \eprint{1307.6554}.

\bibitem[{\citenamefont{Batell et~al.}(2014)\citenamefont{Batell, Essig, and
  Surujon}}]{Batell:2014mga}
\bibinfo{author}{\bibfnamefont{B.}~\bibnamefont{Batell}},
  \bibinfo{author}{\bibfnamefont{R.}~\bibnamefont{Essig}}, \bibnamefont{and}
  \bibinfo{author}{\bibfnamefont{Z.}~\bibnamefont{Surujon}}, ``{Strong
  Constraints on Sub-GeV Dark Sectors from SLAC Beam Dump E137},''
  \bibinfo{journal}{Phys. Rev. Lett.} \textbf{\bibinfo{volume}{113}},
  \bibinfo{pages}{171802} (\bibinfo{year}{2014}), \eprint{1406.2698}.

\bibitem[{\citenamefont{Kahn et~al.}(2015)\citenamefont{Kahn, Krnjaic, Thaler,
  and Toups}}]{Kahn:2014sra}
\bibinfo{author}{\bibfnamefont{Y.}~\bibnamefont{Kahn}},
  \bibinfo{author}{\bibfnamefont{G.}~\bibnamefont{Krnjaic}},
  \bibinfo{author}{\bibfnamefont{J.}~\bibnamefont{Thaler}}, \bibnamefont{and}
  \bibinfo{author}{\bibfnamefont{M.}~\bibnamefont{Toups}}, ``{DAEDALUS and dark
  matter detection},'' \bibinfo{journal}{Phys. Rev.}
  \textbf{\bibinfo{volume}{D91}}, \bibinfo{pages}{055006}
  (\bibinfo{year}{2015}), \eprint{1411.1055}.

\bibitem[{\citenamefont{Dobrescu and Frugiuele}(2015)}]{Dobrescu:2014ita}
\bibinfo{author}{\bibfnamefont{B.~A.} \bibnamefont{Dobrescu}} \bibnamefont{and}
  \bibinfo{author}{\bibfnamefont{C.}~\bibnamefont{Frugiuele}}, ``{GeV-Scale
  Dark Matter: Production at the Main Injector},'' \bibinfo{journal}{JHEP}
  \textbf{\bibinfo{volume}{02}}, \bibinfo{pages}{019} (\bibinfo{year}{2015}),
  \eprint{1410.1566}.

\bibitem[{\citenamefont{Coloma et~al.}(2016)\citenamefont{Coloma, Dobrescu,
  Frugiuele, and Harnik}}]{Coloma:2015pih}
\bibinfo{author}{\bibfnamefont{P.}~\bibnamefont{Coloma}},
  \bibinfo{author}{\bibfnamefont{B.~A.} \bibnamefont{Dobrescu}},
  \bibinfo{author}{\bibfnamefont{C.}~\bibnamefont{Frugiuele}},
  \bibnamefont{and} \bibinfo{author}{\bibfnamefont{R.}~\bibnamefont{Harnik}},
  ``{Dark matter beams at LBNF},'' \bibinfo{journal}{JHEP}
  \textbf{\bibinfo{volume}{04}}, \bibinfo{pages}{047} (\bibinfo{year}{2016}),
  \eprint{1512.03852}.

\bibitem[{\citenamefont{deNiverville et~al.}(2017)\citenamefont{deNiverville,
  Chen, Pospelov, and Ritz}}]{deNiverville:2016rqh}
\bibinfo{author}{\bibfnamefont{P.}~\bibnamefont{deNiverville}},
  \bibinfo{author}{\bibfnamefont{C.-Y.} \bibnamefont{Chen}},
  \bibinfo{author}{\bibfnamefont{M.}~\bibnamefont{Pospelov}}, \bibnamefont{and}
  \bibinfo{author}{\bibfnamefont{A.}~\bibnamefont{Ritz}}, ``{Light dark matter
  in neutrino beams: production modelling and scattering signatures at
  MiniBooNE, T2K and SHiP},'' \bibinfo{journal}{Phys. Rev.}
  \textbf{\bibinfo{volume}{D95}}, \bibinfo{pages}{035006}
  (\bibinfo{year}{2017}), \eprint{1609.01770}.

\bibitem[{\citenamefont{Pospelov and Tsai}(2018)}]{Pospelov:2017kep}
\bibinfo{author}{\bibfnamefont{M.}~\bibnamefont{Pospelov}} \bibnamefont{and}
  \bibinfo{author}{\bibfnamefont{Y.-D.} \bibnamefont{Tsai}}, ``{Light scalars
  and dark photons in Borexino and LSND experiments},'' \bibinfo{journal}{Phys.
  Lett.} \textbf{\bibinfo{volume}{B785}}, \bibinfo{pages}{288}
  (\bibinfo{year}{2018}), \eprint{1706.00424}.

\bibitem[{\citenamefont{Magill et~al.}(2018{\natexlab{b}})\citenamefont{Magill,
  Plestid, Pospelov, and Tsai}}]{Magill:2018jla}
\bibinfo{author}{\bibfnamefont{G.}~\bibnamefont{Magill}},
  \bibinfo{author}{\bibfnamefont{R.}~\bibnamefont{Plestid}},
  \bibinfo{author}{\bibfnamefont{M.}~\bibnamefont{Pospelov}}, \bibnamefont{and}
  \bibinfo{author}{\bibfnamefont{Y.-D.} \bibnamefont{Tsai}}, ``{Dipole portal
  to heavy neutral leptons},''  (\bibinfo{year}{2018}{\natexlab{b}}),
  \eprint{1803.03262}.

\bibitem[{\citenamefont{Bowman et~al.}(2018)\citenamefont{Bowman, Rogers,
  Monsalve, Mozdzen, and Mahesh}}]{Bowman:2018yin}
\bibinfo{author}{\bibfnamefont{J.~D.} \bibnamefont{Bowman}},
  \bibinfo{author}{\bibfnamefont{A.~E.~E.} \bibnamefont{Rogers}},
  \bibinfo{author}{\bibfnamefont{R.~A.} \bibnamefont{Monsalve}},
  \bibinfo{author}{\bibfnamefont{T.~J.} \bibnamefont{Mozdzen}},
  \bibnamefont{and} \bibinfo{author}{\bibfnamefont{N.}~\bibnamefont{Mahesh}},
  ``{An absorption profile centred at 78 megahertz in the sky-averaged
  spectrum},'' \bibinfo{journal}{Nature} \textbf{\bibinfo{volume}{555}},
  \bibinfo{pages}{67} (\bibinfo{year}{2018}).

\bibitem[{\citenamefont{Barkana}(2018)}]{Barkana:2018lgd}
\bibinfo{author}{\bibfnamefont{R.}~\bibnamefont{Barkana}}, ``{Possible
  interaction between baryons and dark-matter particles revealed by the first
  stars},'' \bibinfo{journal}{Nature} \textbf{\bibinfo{volume}{555}},
  \bibinfo{pages}{71} (\bibinfo{year}{2018}), \eprint{1803.06698}.

\bibitem[{\citenamefont{Mu{\~n}oz and Loeb}(2018)}]{Munoz2018}
\bibinfo{author}{\bibfnamefont{J.~B.} \bibnamefont{Mu{\~n}oz}}
  \bibnamefont{and} \bibinfo{author}{\bibfnamefont{A.}~\bibnamefont{Loeb}},
  ``{Insights on Dark Matter from Hydrogen during Cosmic Dawn},''
  (\bibinfo{year}{2018}), \eprint{1802.10094}.

\bibitem[{\citenamefont{Berlin et~al.}(2018{\natexlab{a}})\citenamefont{Berlin,
  Hooper, Krnjaic, and McDermott}}]{Berlin:2018sjs}
\bibinfo{author}{\bibfnamefont{A.}~\bibnamefont{Berlin}},
  \bibinfo{author}{\bibfnamefont{D.}~\bibnamefont{Hooper}},
  \bibinfo{author}{\bibfnamefont{G.}~\bibnamefont{Krnjaic}}, \bibnamefont{and}
  \bibinfo{author}{\bibfnamefont{S.~D.} \bibnamefont{McDermott}}, ``{Severely
  Constraining Dark Matter Interpretations of the 21-cm Anomaly},''
  (\bibinfo{year}{2018}{\natexlab{a}}), \eprint{1803.02804}.

\bibitem[{\citenamefont{Barkana et~al.}(2018)\citenamefont{Barkana,
  Outmezguine, Redigolo, and Volansky}}]{Barkana:2018qrx}
\bibinfo{author}{\bibfnamefont{R.}~\bibnamefont{Barkana}},
  \bibinfo{author}{\bibfnamefont{N.~J.} \bibnamefont{Outmezguine}},
  \bibinfo{author}{\bibfnamefont{D.}~\bibnamefont{Redigolo}}, \bibnamefont{and}
  \bibinfo{author}{\bibfnamefont{T.}~\bibnamefont{Volansky}}, ``{Signs of Dark
  Matter at 21-cm?},''  (\bibinfo{year}{2018}), \eprint{1803.03091}.

\bibitem[{\citenamefont{Slatyer and Wu}(2018)}]{Slatyer:2018aqg}
\bibinfo{author}{\bibfnamefont{T.~R.} \bibnamefont{Slatyer}} \bibnamefont{and}
  \bibinfo{author}{\bibfnamefont{C.-L.} \bibnamefont{Wu}}, ``{Early-Universe
  constraints on dark matter-baryon scattering and their implications for a
  global 21 cm signal},'' \bibinfo{journal}{Phys. Rev.}
  \textbf{\bibinfo{volume}{D98}}, \bibinfo{pages}{023013}
  (\bibinfo{year}{2018}), \eprint{1803.09734}.

\bibitem[{\citenamefont{Liu and Zhang}(2018)}]{Liu:2018jdi}
\bibinfo{author}{\bibfnamefont{Z.}~\bibnamefont{Liu}} \bibnamefont{and}
  \bibinfo{author}{\bibfnamefont{Y.}~\bibnamefont{Zhang}}, ``{Probing
  millicharge at BESIII},''  (\bibinfo{year}{2018}), \eprint{1808.00983}.

\bibitem[{\citenamefont{Berlin et~al.}(2018{\natexlab{b}})\citenamefont{Berlin,
  Blinov, Krnjaic, Schuster, and Toro}}]{Berlin:2018bsc}
\bibinfo{author}{\bibfnamefont{A.}~\bibnamefont{Berlin}},
  \bibinfo{author}{\bibfnamefont{N.}~\bibnamefont{Blinov}},
  \bibinfo{author}{\bibfnamefont{G.}~\bibnamefont{Krnjaic}},
  \bibinfo{author}{\bibfnamefont{P.}~\bibnamefont{Schuster}}, \bibnamefont{and}
  \bibinfo{author}{\bibfnamefont{N.}~\bibnamefont{Toro}}, ``{Dark Matter,
  Millicharges, Axion and Scalar Particles, Gauge Bosons, and Other New Physics
  with LDMX},''  (\bibinfo{year}{2018}{\natexlab{b}}), \eprint{1807.01730}.

\bibitem[{\citenamefont{Gninenko et~al.}(2018)\citenamefont{Gninenko,
  Kirpichnikov, and Krasnikov}}]{Gninenko:2018ter}
\bibinfo{author}{\bibfnamefont{S.~N.} \bibnamefont{Gninenko}},
  \bibinfo{author}{\bibfnamefont{D.~V.} \bibnamefont{Kirpichnikov}},
  \bibnamefont{and} \bibinfo{author}{\bibfnamefont{N.~V.}
  \bibnamefont{Krasnikov}}, ``{Probing millicharged particles with NA64
  experiment at CERN},''  (\bibinfo{year}{2018}), \eprint{1810.06856}.

\bibitem[{\citenamefont{Singh et~al.}(2018)}]{Singh:2018von}
\bibinfo{author}{\bibfnamefont{L.}~\bibnamefont{Singh}} \bibnamefont{et~al.},
  ``{Constraints on millicharged particles with low threshold germanium
  detectors at Kuo-Sheng Reactor Neutrino Laboratory},''
  (\bibinfo{year}{2018}), \eprint{1808.02719}.

\bibitem[{\citenamefont{Tsai}(2018)}]{Tsai:2018_Thesis}
\bibinfo{author}{\bibfnamefont{Y.-D.} \bibnamefont{Tsai}}, Ph.D. thesis,
  \bibinfo{school}{Cornell U., Phys. Dept.} (\bibinfo{year}{2018}).

\bibitem[{MCP()}]{MCP_Proposal}
\bibinfo{note}{Kevin Kelly, Ryan Plestid, Maxim Pospelov, Yu-Dai Tsai, in
  progress.}

\bibitem[{\citenamefont{Izaguirre and Yavin}(2015)}]{Izaguirre:2015eya}
\bibinfo{author}{\bibfnamefont{E.}~\bibnamefont{Izaguirre}} \bibnamefont{and}
  \bibinfo{author}{\bibfnamefont{I.}~\bibnamefont{Yavin}}, ``{New window to
  millicharged particles at the LHC},'' \bibinfo{journal}{Phys. Rev.}
  \textbf{\bibinfo{volume}{D92}}, \bibinfo{pages}{035014}
  (\bibinfo{year}{2015}), \eprint{1506.04760}.

\bibitem[{\citenamefont{{A. Haas}}()}]{milliQan:slides}
\bibinfo{author}{\bibnamefont{{A. Haas}}},
  ``{\url{https://web.fnal.gov/organization/theory/JETP/2016/Haas_milliQan_FermilabWC_9-8-2017.pdf}},''.

\bibitem[{\citenamefont{Sher and Stevens}(2018)}]{Sher:2017wya}
\bibinfo{author}{\bibfnamefont{M.}~\bibnamefont{Sher}} \bibnamefont{and}
  \bibinfo{author}{\bibfnamefont{J.}~\bibnamefont{Stevens}}, ``{Detecting a
  heavy neutrino electric dipole moment at the LHC},'' \bibinfo{journal}{Phys.
  Lett.} \textbf{\bibinfo{volume}{B777}}, \bibinfo{pages}{246}
  (\bibinfo{year}{2018}), \eprint{1710.06894}.

\bibitem[{\citenamefont{Sj{\"o}strand et~al.}(2015)\citenamefont{Sj{\"o}strand,
  Ask, Christiansen, Corke, Desai, Ilten, Mrenna, Prestel, Rasmussen, and
  Skands}}]{Sjostrand:2014zea}
\bibinfo{author}{\bibfnamefont{T.}~\bibnamefont{Sj{\"o}strand}},
  \bibinfo{author}{\bibfnamefont{S.}~\bibnamefont{Ask}},
  \bibinfo{author}{\bibfnamefont{J.~R.} \bibnamefont{Christiansen}},
  \bibinfo{author}{\bibfnamefont{R.}~\bibnamefont{Corke}},
  \bibinfo{author}{\bibfnamefont{N.}~\bibnamefont{Desai}},
  \bibinfo{author}{\bibfnamefont{P.}~\bibnamefont{Ilten}},
  \bibinfo{author}{\bibfnamefont{S.}~\bibnamefont{Mrenna}},
  \bibinfo{author}{\bibfnamefont{S.}~\bibnamefont{Prestel}},
  \bibinfo{author}{\bibfnamefont{C.~O.} \bibnamefont{Rasmussen}},
  \bibnamefont{and} \bibinfo{author}{\bibfnamefont{P.~Z.}
  \bibnamefont{Skands}}, ``{An Introduction to PYTHIA 8.2},''
  \bibinfo{journal}{Comput. Phys. Commun.} \textbf{\bibinfo{volume}{191}},
  \bibinfo{pages}{159} (\bibinfo{year}{2015}), \eprint{1410.3012}.

\bibitem[{\citenamefont{Martin et~al.}(2009)\citenamefont{Martin, Stirling,
  Thorne, and Watt}}]{Martin:2009iq}
\bibinfo{author}{\bibfnamefont{A.~D.} \bibnamefont{Martin}},
  \bibinfo{author}{\bibfnamefont{W.~J.} \bibnamefont{Stirling}},
  \bibinfo{author}{\bibfnamefont{R.~S.} \bibnamefont{Thorne}},
  \bibnamefont{and} \bibinfo{author}{\bibfnamefont{G.}~\bibnamefont{Watt}},
  ``{Parton distributions for the LHC},'' \bibinfo{journal}{Eur. Phys. J.}
  \textbf{\bibinfo{volume}{C63}}, \bibinfo{pages}{189} (\bibinfo{year}{2009}),
  \eprint{0901.0002}.

\bibitem[{\citenamefont{Adamson et~al.}(2016)}]{Adamson:2015dkw}
\bibinfo{author}{\bibfnamefont{P.}~\bibnamefont{Adamson}} \bibnamefont{et~al.},
  ``{The NuMI Neutrino Beam},'' \bibinfo{journal}{Nucl. Instrum. Meth.}
  \textbf{\bibinfo{volume}{A806}}, \bibinfo{pages}{279} (\bibinfo{year}{2016}),
  \eprint{1507.06690}.

\bibitem[{\citenamefont{Acciarri and \emph{et.
  al.}}(2015)}]{DUNECollaboration2015}
\bibinfo{author}{\bibfnamefont{R.}~\bibnamefont{Acciarri}} \bibnamefont{and}
  \bibinfo{author}{\bibnamefont{\emph{et. al.}}}, ``{Long-Baseline Neutrino
  Facility (LBNF) and Deep Underground Neutrino Experiment (DUNE) Conceptual
  Design Report Volume 2: The Physics Program for DUNE at LBNF},''
  \textbf{\bibinfo{volume}{2}} (\bibinfo{year}{2015}), \eprint{1512.06148},
  \urlprefix\url{http://arxiv.org/abs/1512.06148}.

\bibitem[{\citenamefont{Patrignani et~al.}(2016)}]{Patrignani:2016xqp}
\bibinfo{author}{\bibfnamefont{C.}~\bibnamefont{Patrignani}}
  \bibnamefont{et~al.} (\bibinfo{collaboration}{Particle Data Group}),
  ``{Review of Particle Physics},'' \bibinfo{journal}{Chin. Phys.}
  \textbf{\bibinfo{volume}{C40}}, \bibinfo{pages}{100001}
  (\bibinfo{year}{2016}).

\bibitem[{\citenamefont{Tanabashi et~al.}(2018)}]{Tanabashi:2018oca}
\bibinfo{author}{\bibfnamefont{M.}~\bibnamefont{Tanabashi}}
  \bibnamefont{et~al.} (\bibinfo{collaboration}{Particle Data Group}),
  ``{Review of Particle Physics},'' \bibinfo{journal}{Phys. Rev.}
  \textbf{\bibinfo{volume}{D98}}, \bibinfo{pages}{030001}
  (\bibinfo{year}{2018}).

\bibitem[{\citenamefont{Birks}(1951)}]{Birks:1951boa}
\bibinfo{author}{\bibfnamefont{J.~B.} \bibnamefont{Birks}}, ``{Scintillations
  from Organic Crystals: Specific Fluorescence and Relative Response to
  Different Radiations},'' \bibinfo{journal}{Proc. Phys. Soc.}
  \textbf{\bibinfo{volume}{A64}}, \bibinfo{pages}{874} (\bibinfo{year}{1951}).

\bibitem[{\citenamefont{{J. Conrad}}()}]{wunderbar1}
\bibinfo{author}{\bibnamefont{{J. Conrad}}},
  ``{\url{https://indico.fnal.gov/event/11326/contribution/1/material/slides/0.pdf}},''.

\bibitem[{\citenamefont{{D. Whittington}}()}]{wunderbar2}
\bibinfo{author}{\bibnamefont{{D. Whittington}}},
  ``{\url{https://indico.fnal.gov/event/12089/session/3/material/slides/0?contribId=11}},''.

\bibitem[{Pre()}]{Precite_Roni}
\bibinfo{note}{R. Harnik, Z. Liu, O. Palamara, in progress.}

\bibitem[{\citenamefont{{R. Harnik}}()}]{Roni:slides}
\bibinfo{author}{\bibnamefont{{R. Harnik}}},
  ``{\url{https://indico.fnal.gov/event/18430/session/8/contribution/17/material/slides/0.pdf}},''.

\bibitem[{\citenamefont{{J. Cai et al.}}()}]{duneprism}
\bibinfo{author}{\bibnamefont{{J. Cai et al.}}},
  ``{\url{https://indico.fnal.gov/event/16205/contribution/2/material/0/0.pdf}},''.

\bibitem[{\citenamefont{Aguilar-Arevalo
  et~al.}(2018)}]{Aguilar-Arevalo:2018wea}
\bibinfo{author}{\bibfnamefont{A.~A.} \bibnamefont{Aguilar-Arevalo}}
  \bibnamefont{et~al.} (\bibinfo{collaboration}{MiniBooNE DM}), ``{Dark Matter
  Search in Nucleon, Pion, and Electron Channels from a Proton Beam Dump with
  MiniBooNE},''  (\bibinfo{year}{2018}), \eprint{1807.06137}.

\bibitem[{\citenamefont{Miramonti}(2005)}]{Miramonti:2005xq}
\bibinfo{author}{\bibfnamefont{L.}~\bibnamefont{Miramonti}}, ``{European
  underground laboratories: An Overview},'' \bibinfo{journal}{AIP Conf. Proc.}
  \textbf{\bibinfo{volume}{785}}, \bibinfo{pages}{3} (\bibinfo{year}{2005}),
  \bibinfo{note}{[,3(2005)]}, \eprint{hep-ex/0503054}.

\bibitem[{\citenamefont{Acciarri et~al.}(2018)}]{Acciarri:2018myr}
\bibinfo{author}{\bibfnamefont{R.}~\bibnamefont{Acciarri}} \bibnamefont{et~al.}
  (\bibinfo{collaboration}{ArgoNeuT}), ``{Demonstration of MeV-Scale Physics in
  Liquid Argon Time Projection Chambers Using ArgoNeuT},''
  (\bibinfo{year}{2018}), \eprint{1810.06502}.

\bibitem[{\citenamefont{Cortina~Gil et~al.}(2017)}]{NA62:2017rwk}
\bibinfo{author}{\bibfnamefont{E.}~\bibnamefont{Cortina~Gil}}
  \bibnamefont{et~al.} (\bibinfo{collaboration}{NA62}), ``{The Beam and
  detector of the NA62 experiment at CERN},'' \bibinfo{journal}{JINST}
  \textbf{\bibinfo{volume}{12}}, \bibinfo{pages}{P05025}
  (\bibinfo{year}{2017}), \eprint{1703.08501}.

\bibitem[{\citenamefont{Ahdida et~al.}(2018)}]{SHiP:2018yqc}
\bibinfo{author}{\bibfnamefont{C.}~\bibnamefont{Ahdida}} \bibnamefont{et~al.}
  (\bibinfo{collaboration}{SHiP}), ``{The experimental facility for the Search
  for Hidden Particles at the CERN SPS},''  (\bibinfo{year}{2018}),
  \eprint{1810.06880}.

\bibitem[{\citenamefont{Yoo}(2018)}]{Yoo:2018lhk}
\bibinfo{author}{\bibfnamefont{J.~H.} \bibnamefont{Yoo}}
  (\bibinfo{collaboration}{milliQan}), ``{The milliQan Experiment: Search for
  milli-charged Particles at the LHC},''  (\bibinfo{year}{2018}),
  \eprint{1810.06733}.

\end{thebibliography}
\bibliographystyle{apsrev-title}

\end{document}